\documentclass[twocolumn]{aa}
\usepackage{graphicx}
\usepackage{txfonts}
 \usepackage{natbib}
  \bibpunct{(}{)}{;}{a}{}{,} 

\newcommand{\OB}{[OIII]$\lambda$5007/H$\beta$}
\newcommand{\NA}{[NII]$\lambda$6583/H$\alpha$}
\newcommand{\SA}{[SII]$\lambda$6717+6731/H$\alpha$}

\begin{document}
   \title{The merging/AGN connection}
   \subtitle{II. Ionization of the circumnuclear regions}

\author{S.F.S\'anchez\inst{1}\thanks{Based on observations made with ESO
    Telescopes at the Paranal Observatory under programme ID
    71.B-3055(A)}~\thanks{Visiting Astronomer,
    German-Spanish Astronomical Centre of Calar Alto.
}
\and T.Becker\inst{1} 
\and  B.Garcia-Lorenzo\inst{2} 
\and C.R.Benn\inst{3}
 \and L.Christensen\inst{1}
\and A.Kelz\inst{1} 
    \and K.Jahnke\inst{1} 
 \and M.M.Roth \inst{1}}

   \offprints{S.F.S\'anchez: ssanchez@aip.de}

\institute{Astrophysikalisches Institut Potsdam, And der Sternwarte 16, 14482
  Potsdam, Germany 
\and 
Instituto de Astrof\'\i sica de Canarias, 38205 La Laguna, Tenerife, Spain 
\and Isaac Newton Group, Apt. 321, 38700 Santa Cruz de La Palma, Spain }

   \date{;}

   \abstract{We report the first results of a study of a sample of 20
     galaxy mergers/interacting systems, using the VIMOS and PMAS
    integral field spectrographs.
For each object, we extracted the integrated spectrum of the
     central regions
and analyzed the ionization state using classical diagnostic diagrams
   \citep{veil87}. There is evidence of AGN ionization in 4 of the objects,
   i.e. 20\% of the sample, a considerably higher fraction
than
  found in previous studies ($\sim$4\%).
An increase of the nuclear activity with the far infrared temperature may
account for the difference.
   }

   \maketitle
%

\section{Introduction}

Interactions and mergers can strongly affect galaxy evolution.  Since the
early studies by \cite{toom72} and \cite{lars78}, it is generally believed
that interactions can trigger starburst processes in galaxies, via tidal
effects. The fact that interacting galaxies are more luminous in the far
infrared (FIR) \citep[e.g.][]{sand96} and the radio \citep[e.g.][]{humm81},
and that they have larger H$\alpha$ equivalent widths \citep[e.g.][]{kenn87}
supports this hypothesis. The star formation is found to be stronger in the
nucleus, although it is also enhanced in the disk.  A widely accepted idea,
supported by simulations, is that galaxy interactions induce flows of gas from
the outer parts of the galaxy and/or the companion into the inner regions
through loss of the angular momentum induced by tidal forces
\citep[e.g.][]{miho96}.

It has long been suggested that strong interactions and mergers may trigger
nuclear activity. Many authors have found that quasar host galaxies show
distorted morphologies, reminiscent of past merging events
\citep[e.g.][]{mcle94a,mcle94b,bahc97,sanc03b,jahn04}. However, other authors found
no evidence of past interactions \citep[e.g.][]{dunl03}.  It is not clear if
the fraction of AGN host galaxies undergoing interactions is really larger
than the fraction of inactive galaxies undergoing interactions, or how the
timescales of the morphological relaxation and the AGN ignition relate
\citep[e.g.][]{sanc04c}.  However, it is clear that some AGN, if not all, are
generated by mergers/interactions between galaxies \citep{cana01,catt04}.

The fraction of AGNs in merging systems increases with the FIR luminosity,
being $\sim$30\% for the extreme case of the ultraluminous infrared galaxies
(ULIRGs) \citep{bush02}, with FIR luminosity $L_{FIR}>10^{12} L_\odot$.
\cite{sand88} suggested an evolutionary scenario, in which major mergers may
allow the infall of gas towards the nuclear regions, triggering nuclear star
formation and igniting an AGN. Both the AGN and the star formation heat the
dust, and the object is observed as a ULIRG.  \cite{coli01} proposed a
modification of this scenario in which cool ULIRGs would be generated during
the merging of two or more low-mass sub-$L^*$ spirals, and warm ULIRGs in the
merging of an $L^*$ galaxy with a substantially less massive galaxy or during
the merging of two intermediate-mass spirals. Under this hypothesis only warm
ULIRGs would evolve/host an AGN, while cool ULIRGs would be heated by star
formation.

At lower luminosities, $L_{FIR}\sim 10^{10-11} L_\odot$, there is no evidence
of an excess of AGNs among interacting/merging systems compared to isolated
galaxies \cite[e.g.][]{daha85, keel85, bush88, seki92, liu95, don99, donz00,
  berg03}. This deficiency could be partially explained by the presence of
strong circumnuclear star formation that can blur the AGN signatures
\citep{don99,donz00}. There is significantly more nuclear star formation in
interacting galaxies, compared to isolated ones \cite[e.g.][]{berg03}. This
nuclear star formation increases with the FIR temperature
\citep{bush88,berg03}.  If the scenario proposed by \cite{coli01} is valid for
lower-luminosity objects, the nuclear SFR and the probability of generating an
AGN could be larger in warm FIR sources than in cool ones.  \cite{grij85} and
\cite{broe91} already noticed an increase of the number of AGNs with the FIR
temperature, using slit spectroscopy.

Most previous studies of this topic were based on samples of
interacting/merging galaxies selected by morphology
\citep[e.g.][]{donz97,berg03}, without considering the FIR properties.  As a
consequence, these samples contain mainly cool FIR sources (log$_{10}
(f_{60\mu}/f_{100\mu})\sim-0.4$). We selected a sample of 20 IRAS sources with
evidence of interactions/major merging \citep[][, paper I]{sanc04a}, with
similar FIR luminosities to previous samples, $L_{FIR}\sim 10^{10-11} L_\odot$,
but warmer (log$_{10}(f_{60\mu}/f_{100\mu})\sim-0.25$).  Since the use of slit
spectroscopy can bias the results (paper I), we obtained integral-field
spectra of all them to explore the connection between galaxy interactions and
nuclear activity.

\section{Observations and data reduction }
\label{obs}

The spectra were taken with the integral-field units (IFUs) of VIMOS
\citep{lefe03}, 17 objects, and PMAS \citep{roth00}, 3 objects.  The VIMOS
observations were made on the nights of the 9th, 10th, 22th, 23th, 24th and
26th of April 2003 at the VLT/UT3,
in clear sky conditions, 
with an average seeing $\sim$1.6$\arcsec$. We obtained, for each object, 3
exposures of 300s using each of the low-resolution red (LRr) and blue (LRb)
grisms.  With this setup we cover the wavelength ranges 6800-9500 \AA\ (LRr)
and 3900-6800 \AA\ (LRb) with resolutions of 7.14 \AA/pixel (LRr) and 5.35
\AA/pixel (LRb), respectively. The VIMOS IFU contains a focal-plane array of
6400 lenslets each of size 0.67$\arcsec$$\times$0.67$\arcsec$ (in the
low-resolution mode) distributed on a regular grid. The final field-of-view
was 54$\arcsec$$\times$54$\arcsec$, which allows us to cover the different
components within the IRAS aperture ($\sim$1$\arcmin$). In two cases (IRAS
10219-2828 and IRAS 12110-3412) we obtained data for only one grism (LRr and
LRb, respectively).

\begin{figure}
   \centering
\resizebox{\hsize}{!}
{\includegraphics[width=\hsize]{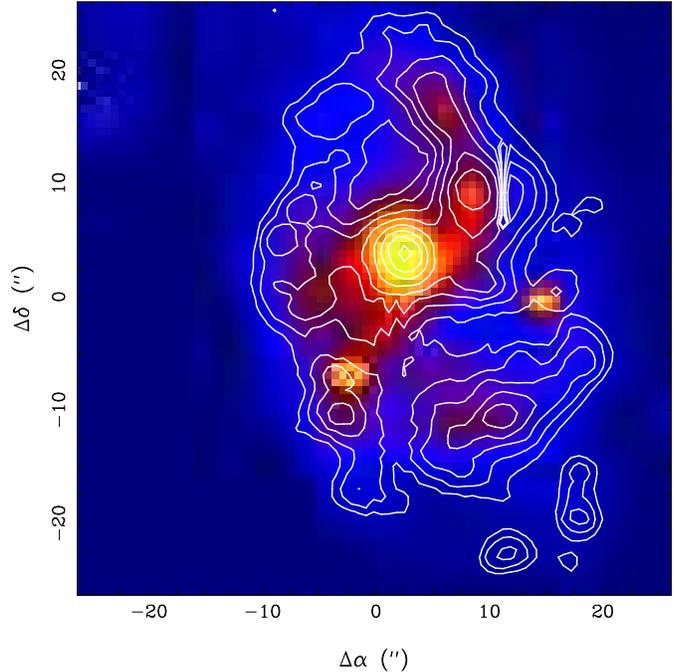}}
      \caption{ 
Intensity map of the continnum emission of IRAS
13477-4848 at wavelengths $\sim$6477-6591 \AA,
extracted from the VIMOS datacube. The contours show the emission
in H$\alpha$+[NII] on a logarithm scale, obtained by subtracting the
continuum emission from that at wavelengths centred on 
the emission lines
($\sim$6612-6670 \AA). }
         \label{eg}
   \end{figure}

The PMAS data were taken during the nights of the 28th of April, and 5th and
8th of August 2003 at the 3.5-m telescope of the Calar Alto Observatory.  The
observations were made in poor weather conditions as a back-up project for
these nights. We used the V300 grating which covers the wavelength range
$\sim$4000 - 7700 \AA~ with a resolution of 3.3 \AA/pix.  The exact wavelength
range depends on the rotation angle selected for the grating, which changes
from object to object.  The PMAS focal plane contains an array of 256 lenslets
each of size 0.5$\arcsec$$\times$0.5$\arcsec$ (in the standard configuration)
distributed on a square grid. The field-of-view is
8$\arcsec$$\times$8$\arcsec$, i.e. too small to cover the IRAS aperture with a
single pointing.  Different mosaic patterns were used for each object to cover
the central regions and different regions of interest (e.g., tidal tails). Due
to the variable atmospheric conditions different exposure times were used for
each object and each night, to obtain similar signal-to-noise levels in all
the frames. The exposure times ranged from 1200 s per pointing for IRAS
16365+4202 (05/08/03) to 5400 s for IRAS 12110+1624 (28/04/03).

Both the VIMOS and PMAS data were reduced using the P3d reduction package,
written in IDL \citep{beck02}. Although initially written to reduce PMAS data,
it was slightly modified to reduce data from other IFUs, such as VIMOS
\cite[e.g.][]{sanc04b}. 
The expected locations of the spectra were traced on a continuum-lamp exposure
obtained before each target exposure.  After bias subtraction, the target
spectra were extracted by adding the signal from the 5 pixels around the
central traced pixel. The continuum-lamp exposure was also used to determine
the fiber-to-fiber response of the instrument at each particular wavelength
(the so-called flat-spectra). These flat-spectra were used to homogenize the
response of all the fibers. Wavelength calibration was carried out via
observation of emission-line lamps, also obtained before the science
exposures.  The accuracy of the wavelength calibration was checked using sky
emission lines, and has standard deviation 1.2 \AA~(VIMOS-LRr), 1.0
\AA~(VIMOS-LRb) and 0.5 \AA~(PMAS).  Different exposures taken at the same
pointing were combined using IRAF tasks. A 3$\sigma$ clipping algorithm was
used to remove possible cosmic rays. The different PMAS pointings were
mosaiced into a single frame, using the Euro3D data format, as described in
\citep{sanc04b}.

Further analysis was carried out using E3D \citep{sanc04}, and our own fitting
routines (S\'anchez et al., in preparation). An average sky spectrum, taken
from areas uncontaminated by target spectra, was subtracted from the target
spectra.


\section{Analysis and results}
\label{ana}

\subsection{Extraction of the nuclear spectra}

Figure \ref{eg} shows an example of the reduced data: a 
continuum-intensity 
map of IRAS 13477-4848 at wavelengths $\sim$6477-6591 \AA, 
extracted from the VIMOS datacube.  For each object we extracted
the spectrum of the central region, by co-adding the spectra within a box
aperture of size 2$\arcsec$$\times$2$\arcsec$ (approximately 
the central kiloparsec,
at the redshift of these objects). Whenever the intensity peak of the
continuum emission was not coincident with that of the emission line, we
centered our aperture on the latter to increase the signal-to-noise ratio.
In most cases both peaks were coincident within a couple of arcsec.  This is
illustrated in Fig.  \ref{eg}, where the H$\alpha$+[NII] intensity 
contours are
superimposed on a continuum image.

Six of the 20 objects have two well-separated components.
 with two clear
intensity peaks. The spectra of four of them show emission lines in both
components.
Emission lines are absent in the remaining two. The extracted spectra of
the nuclear regions of the two components have similar characteristics.
Hereafter we restrict our discussion to
the properties of the spectrum of the
brighter component.

\subsection{Ionization Mechanism}

The final dataset consists of the spectra of the nuclear regions of 20
interacting/merging galaxies. Of these, 18 cover a wavelength range
which includes some of the most important emission lines for ionization
diagnostics: i.e., H$\beta$, [OIII]$\lambda$4959, [OIII]$\lambda$5007,
H$\alpha$, [NII]$\lambda$6584, , [NII]$\lambda$6548, [SII]$\lambda$6717 and
[SII]$\lambda$6731. A variety of other emission lines were detected in many of
these spectra. A detailed description of the spectrum of each object will be
presented elsewhere (Sanchez et al., in preparation). 
We fitted the line profiles
to derive the integrated flux of each emission
line, using our own software. A single gaussian was fitted to each 
line, using a polynomial function to characterize
the continuum.  The H$\alpha$-[NII] system and the [SII] doublet were blended
in the LRr VIMOS data. For proper deblending, the minimun line-width was
fixed to be the width measured in isolated sky emission lines (i.e., the
instrumental width).  Our software allows definition
of  emission line systems,
i.e., a kinematically coupled set of emission lines with the same width and
heliocentric velocity. This was essential for accurate deblending
of the lines.
After deriving the flux of each line, the \OB, \NA~ and \SA~ line
ratios were calculated.

\begin{figure}
   \centering
\resizebox{\hsize}{!}
{\includegraphics[width=\hsize,angle=-90]{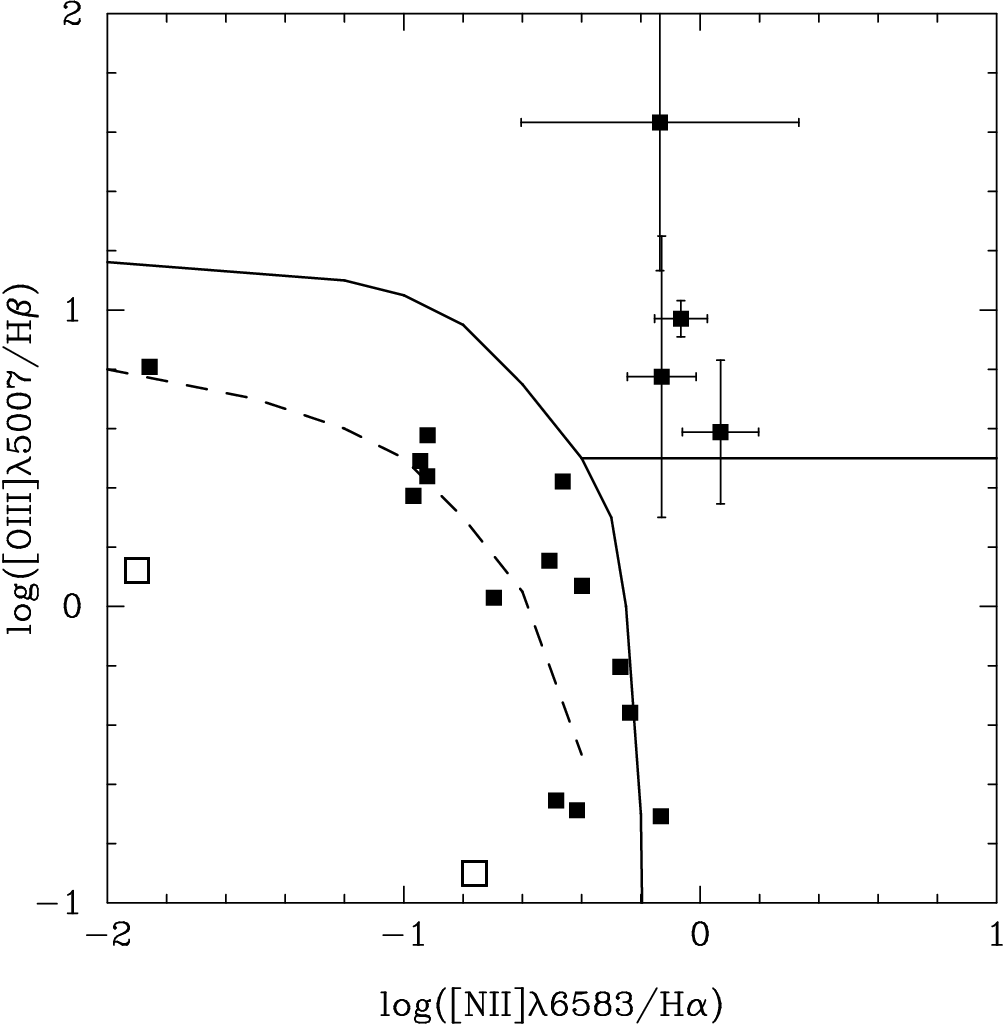}}
      \caption{ \OB~line ratio as a function of \NA~ for the 18 objects 
with both ratios measured
        (solid squares). The open squares indicate the two
        objects 
        with data only for one
        VIMOS grism (LR-red or LR-blue). 
        They have been plotted with an arbitrary value of the
        un-measured line ratio (i.e. at the edge of the figure). 
        The solid lines
        show the separation between ionization due to star formation
        (bottom-left), AGN (top-right) and LINERs (bottom-right).  
        The approximate locus of star-forming objects reported by 
\cite{veil87}, is
         indicated with a dashed line.  }
         \label{diag}
   \end{figure}

   Figure \ref{diag} shows the basic diagnostic diagram, \OB~ vs. \NA~
   \citep{veil87}, for the 18 objects observed with spectral information in
   both wavelength regions (solid squares). The open squares near the edge of
   the plot represent two objects (IRAS 10219-2828 and IRAS 12110-3412) with
   data only for one VIMOS grism (LR-red or LR-blue).  The solid lines show
   the classical separation between ionization due to star formation
   (bottom-left), AGNs (top-right) and LINERs (bottom-right). The boundary
   curve between AGNs and star forming regions is taken from \cite{veil87}. A
   separation between AGNs and LINERs at a ratio of [OIII]/H$\beta = 3$ was
   assumed, following earlier results in the literature
   \citep[e.g.][]{donz00,garcia01}.
   
   The majority of the objects in our sample (14 out of 20) have nuclear
   emission features characteristic of star forming regions. However, for a
   significant fraction of the objects (4 out of 20, 20\%), the nuclear
   emission is dominated by an AGN: IRAS 16330-6820, IRAS 12193-3942, IRAS
   13031-5717 and IRAS 16365+4202. We include the 1$\sigma$ error bars for
   these objects in Fig. 2.  As a check, we repeated the analysis using the
   \OB~vs. \SA~diagnostic diagram. We find the same distribution, reinforcing
   the result from Fig 2.  As expected from ionization models
   \citep[e.g.][]{oste89}, there is a correlation between the \NA~and the
   \SA~line ratios in our sample, with a dispersion dominated by measurement
   errors. The errors in the \NA~ratio are dominated by residual errors in the
   line-deblending procedure, while those of \SA~ (less affected by
   deblending) are dominated by the lower signal-to-noise ratio of
   [SII]$\lambda$6717+6731.


For the two objects with only one 
diagnostic line ratio measured, 
the nature of the ionization is uncertain ,
although one of them, IRAS 12110-3412,
is clearly not an AGN.



\section{Discussion and Conclusions}
\label{disc}

As noted above, most studies of the ionization mechanisms in the
nuclear regions of merging/interacting galaxies are based on studies of cool
FIR sources. These studies
found a small fraction of objects with AGN-like
ionization, typically $\sim$4\% \cite[e.g.][]{don99,berg03}. 
By contrast,
we find a significantly higher fraction of AGNs ($\sim$20\%), in a 
sample of warmer FIR merging/interacting galaxies. This may indicate that the
scenario proposed by \cite{coli01} (that only warm ones host an AGN),
is valid also at lower IR luminosities $L_{FIR}$.

Our current sample corresponds to only $\sim$6\% of all interacting systems
with measured redshift in the IRAS point-source sample \citep{sanc04a}.  How
general is our result? To address this question we compared our sample with
that of \cite{berg03}.  As already mentioned, the two samples cover the same
range of FIR luminosities (indistinguishable, according to a
Kolmogorov-Smirnov test).  However, the ranges of FIR temperatures are
different, with our sample being warmer (the null hyphothesis, that both
samples were drawn from the same parent population, can be rejected at the
0.13~\% level).  If we add the objects from \cite{berg03} within the IRAS
point source catalogue \citep{IRAS94} to our sample, we end up with 77
objects.  Splitting this sample into cool ($f_{60\mu}/f_{100\mu}<-0.3$) and
warm ($f_{60\mu}/f_{100\mu}>-0.3$) FIR sources, we found that only 4\% (2 out
of 50)  of the cool objects harbour an AGN, in compared with 15\% (4 out of
27) of the warm ones.  That is, we still find that the fraction of AGN
increases with the FIR temperature.  We found similar results when comparing
with other samples \cite[e.g.][]{donz97,donz00}.

The probability of igniting an AGN seems to be higher in warm FIR sources,
over a large range of FIR luminosities. According to \cite{coli01}, these
objects are the result of a merger either between a large galaxy and a small
galaxy or between two galaxies of similar size. A possible explanation of this
result could be that apart from the fuel supply provided by the
interaction/merging, the presence of a massive black-hole is required to
ignite an AGN. Massive black-holes are found in massive galaxies, due to the
black-hole/bulge-mass relation \citep[e.g.,][]{mago98,korm01}.
Therefore, only galaxy mergers with one progenitor harbouring a massive
black hole (i.e. a massive galaxy) may (re-)ignite an AGN, and those are the
mergers that generate a warm FIR source.

\begin{acknowledgements}
  
  We thank Dr. E. Laurikainen and Dr. N. Bergvall for providing us with their
  dataset. We thank Dr. A. Stockton, who refereed this article, and help us to
  improve it. This project is founded by the Euro3D RTN on IFS, funded by the
  EC under contract No.  HPRN-CT-2002-00305.  L.Christensen acknowledges
  support by the German Verbundforshung associated with the ULTROS project,
  grant no. 05AE2BAA/4.

\end{acknowledgements}


\end{document}